# Fractal Dimension and Retinal Pathology: A Meta-analysis

Sam Yu[1] and Vasudevan Lakshminarayanan[1,2,3,4]

[1] Department of Physics and Astronomy, University of Waterloo, ON N2L 3G1, Canada; s249yu@uwaterloo.ca (S.Y.)

[2] Theoretical and Experimental Epistemology Lab, School of Optometry and Vision Science; vengulak@uwaterloo.ca (V.L.)

[3] Department of Electrical and Computer Engineering, University of Waterloo, ON N2L 3G1, Canada

[4] Department of Systems Design Engineering, University of Waterloo, ON N2L 3G1, Canada

**Abstract:** Due to the fractal nature of retinal blood vessels, the retinal fractal dimension is a natural parameter for researchers to explore and has garnered interest as a potential diagnostic tool. This review aims to summarize the current scientific evidence regarding the relationship between fractal dimension and retinal pathology and thus assess the clinical value of retinal fractal dimension. Following the PRISMA guidelines, a literature search for research articles was conducted in several internet databases (Embase, PubMed, Web of Science, Scopus). This led to a result of 28 studies included in the final review, which were analyzed via meta-analysis to determine whether the fractal dimension changes significantly in retinal disease versus normal individuals.

## 1. Introduction

*1.1 Retinal Vasculature and Fractal Dimension*

The retina is of crucial importance to eye care professionals as retinal diseases are the leading cause of blindness worldwide. The retina is a thin, light-sensitive neural layer and is supplied by a sophisticated microvascular network that delivers nutrients and carries away waste. As part of the human circulatory system, the network's development tends to seek configurations which minimize operational energy expenditure, reflected by Murray's Law of Minimal Work which relates the radii of parent and daughter vessels, giving rise to the network's branching pattern [1]. Often diseases will have a vascular component that can manifest as abnormalities in this network and thus the network can be observed to acquire insight into the presence(or absence) of disease [2–5]. With advancements in non-invasive ocular imaging techniques such as optical coherence tomography angiography (OCTA) permitting the segmentation of the vasculature into well-defined layers [6], the retinal vasculature has become more accessible than ever for researchers. Hence, increasing attention has been paid to analyzing its quantitative characteristics as a potential diagnostic tool.

The retinal blood vessels form a complex branching pattern that has been shown to be fractal [5]. Therefore, a natural parameter for describing the retinal vasculature is the fractal dimension, first described by [7] and then introduced into ophthalmology by [4]. The fractal dimension is a real number that describes how an object's detail changes at different magnifications. It can be thought of as an extension of the familiar Euclidean dimensions to allow for intermediate states. The fractal dimension of the retinal vascular tree lies between 1 and 2 [5], indicating that its branching pattern fills space more thoroughly than a line, but less than a plane. Thus, the retinal fractal dimension provides a measure of the tree's global branching complexity, which can be altered by the rarefaction or proliferation of blood vessels in a disease scenario. In healthy human subjects, the retinal FD is around 1.7, which is similar to that of a 2D diffusion-limited aggregation process [4,5]. It has been postulated that this is because the retinal vasculature grows through diffusion of angiogenic factors in the retinal plane [8].

*1.2 Common Methods for Measuring Retinal Fractal Dimension*

After image acquisition, retinal images need to be processed first to extract the features of the vascular tree, a process known as *vessel segmentation*, before image binarization and then fractal analysis. Several algorithms exist for this purpose and the reader is referred to the literature [9–11]. We will now briefly discuss commonly used methods for calculating retinal fractal dimension.

*Box-counting (capacity) dimension:* The simplest and most common method used in the literature is the box-counting method [7] for fractal dimension. Given a binarized image of the retinal vascular tree, we overlay the image with a grid of boxes of side-length ε and count how many boxes contain a part of the tree. By decreasing ε, we capture more and more fine details of the tree from the covering. Taking $N(\varepsilon)$ to be the box-count as a function of ε, the box-counting (capacity) dimension [12] is defined as

$$D_{box} = \lim_{\varepsilon \to 0} \frac{\log N(\varepsilon)}{\log 1/\varepsilon} \tag{1}$$

It should be noted that the base of the logarithm does not affect the calculated value.

*Information dimension:* Similar to the method for determining the box-counting dimension, we overlay the retinal image with a grid of boxes of side-length ε. Instead of a box count however, we assign to each box a weight based on its contribution to the tree's information entropy and sum up the weights for each box, defining the information dimension [12,13] as

$$D_{info} = \lim_{\varepsilon \to 0} \frac{\sum_{i=1}^{N(\varepsilon)} p_i \log p_i}{\log 1/\varepsilon} \tag{2}$$

where $N(\varepsilon)$ is the total number of boxes that contain a part of the tree and $p_i = m_i/M$ is the proportion of retinal tree contained in the i-th box; $m_i$ is the number of pixels in the i-th box and $M$ is the total number of pixels in the tree.

*Correlation dimension:* We overlay the retinal image with a grid of side-length ε boxes and define the correlation integral $C(\varepsilon)$ [12] as

$$C(\varepsilon) = \frac{1}{N(\varepsilon)^2} \sum_{i=1, j=1, i \neq j}^{N(\varepsilon)} H(\varepsilon - \|\mathbf{x_i} - \mathbf{x_j}\|) \approx \sum_{i=1}^{N(\varepsilon)} p_i^2, \tag{3}$$

where $H$ is the Heaviside step function and counts the number of tree pixel pairs such that the distance between them is less than ε. The correlation dimension is then defined as

$$D_{corr} = \lim_{\varepsilon \to 0} \frac{\log C(\varepsilon)}{\log (\varepsilon)} \tag{4}$$

*Generalized dimensions:* When discussing multifractal structures such as the retinal vasculature, they are more accurately described by an infinite hierarchy of fractal dimensions [14]. For any real q, the generalized fractal dimension is defined as

$$D_q = \frac{1}{1-q} \lim_{\varepsilon \to 0} \frac{\log I(q, \varepsilon)}{\log 1/\varepsilon} \tag{5}$$

where

$$I(q, \varepsilon) = \sum_{i=1}^{N(\varepsilon)} p_i^q \qquad (6)$$

It can be seen and verified mathematically that $D_0, D_1, D_2$ correspond to the capacity, information, and correlation dimensions respectively as previously discussed. If $q_1 > q_2$ then $D_{q_1} \leq D_{q_2}$ [15], so we have that $D_0 \geq D_1 \geq D_2$, which is a useful check when computing fractal dimensions in practice.

*1.3 Objective*

This objective of this review is to provide an overview of the current scientific evidence of the association between human retinal FD and common retinal disorders, such as diabetic retinopathy (DR), retinal detachment, glaucoma, etc..

*1.4 Research question*

Does the fractal dimension of the retinal vasculature change significantly due to retinal disease when compared to normals?

## 2. Methods

*2.1 Search Strategy*

The Preferred Items for Systematic Review and Meta-Analysis (PRISMA) guidelines [16] were followed for this review. A search was conducted in the databases: EMBASE, MEDLINE, Web of Science, and Scopus. Search queries were constructed by combining relevant subject headings and keywords with Boolean operators.

*2.2 Inclusion/Exclusion Criteria*

The goal of this review is to examine the scientific evidence comprehensively for a wide variety of retinal disorders rather than a select few. The eligibility criteria was determined using the PICO framework [17]. The study populations comprised of participants of any age, race, or gender that had been diagnosed with a retinal disorder. Interventions consisted of calculation of fractal dimensions on processed retinal vasculature images of subjects. Preferably, studies would state the model of the imaging device used, the region of interest which fractal analysis is conducted over, fractal analysis software packages used, and the types of fractal parameters that were calculated. Finally, studies had to compare fractal dimension in case subjects versus control subjects or contrast fractal dimension with disease progression from baseline to follow-ups.

Case-control studies, cross-sectional studies, cohort studies, and case series were included. Studies that employed the following imaging techniques were acceptable:

- Digital fundus photography
- Fundus fluorescein angiography
- Optical coherence tomography angiography (OCTA)
- Scanning laser ophthalmoscopy (SLO)

Reviews and other types of articles without original research were not included. No restrictions on submission date or language were placed.

*2.3 Screening Process*

Search results from each of the databases were exported to Mendeley Reference Manager (Mendeley, London, UK) where duplicates were identified and removed. Subsequently, articles were screened based on title and abstract for relevance.

*2.4 Meta-analysis*

To answer the question of whether the retinal fractal dimension changes significantly due to retinal disease, a meta-analysis was conducted using the data from the included studies. For each disease category, sample mean (SD) fractal dimensions were collected from each study of that category and pooled together to compare normal with case subjects. If the study did not provide sample mean (SD) fractal dimension, then it was estimated from other statistics such as median (IQR) fractal dimension using methods by [18,19]. Otherwise, if there was too much discrepancy, then the study was excluded from the meta-analysis. Mean (SD) fractal dimensions were used to determine effect sizes for each study, which were then used to derive a summary effect size for each disease category using a random-effects model [20].

**3. Results**

*3.1 Results of Study Selection*

A total of 280 records were found initially from the search, leaving 139 records after duplicates were removed. The titles and abstracts were then screened for relevance based off the PICO criteria described in the Methods section. This left 54 studies remaining to be considered, which were all read in-depth, resulting in 28 studies included in the final synthesis with reasons for exclusion for those that did not make the cut. The selection process is summarized in **Figure 1**. 27 of the selected studies exclusively focused on one disease: 10 on diabetic retinopathy (DR), 5 on myopia, 5 on diabetes mellitus (DM) in general, 2 on glaucoma, 2 on hypertension, 1 on macular telangiectasia (MacTel), 1 on retinal occlusions, and 1 on nonarteritic anterior ischemic optic neuropathy (NAION). Only one study examined multiple diseases: DM, hypertension, age-related macular degeneration (AMD), myopia, and glaucoma.

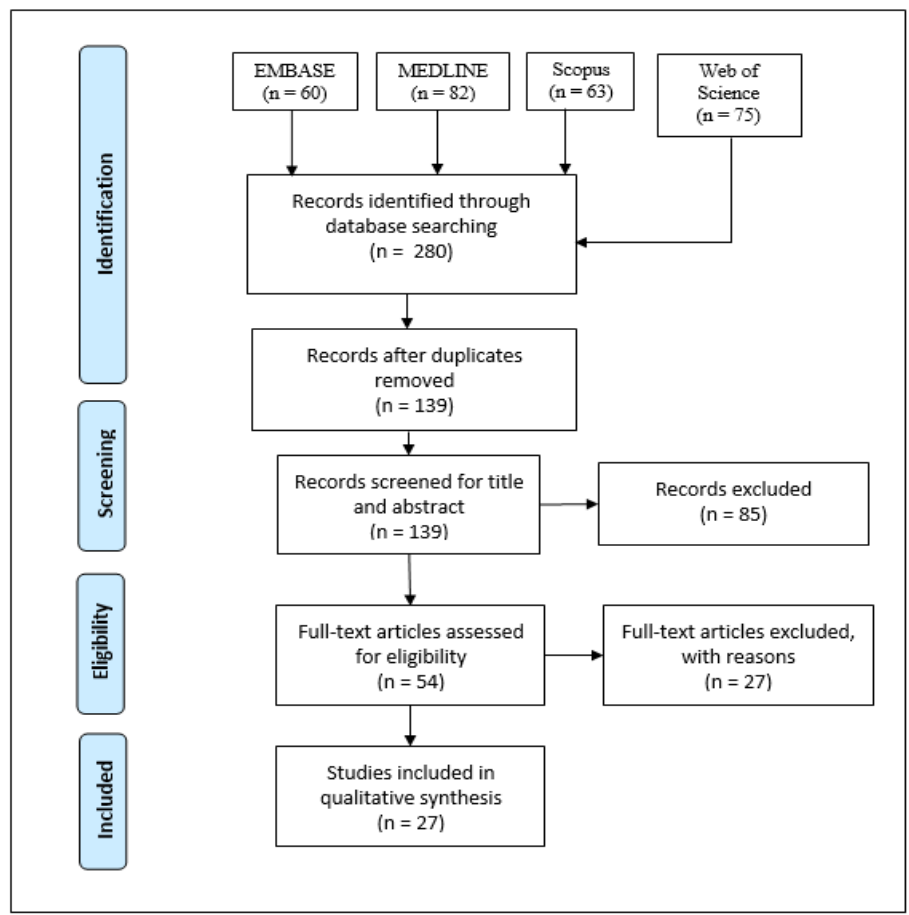

**Figure 1:** The PRISMA flow diagram

1  *3.2 Results of Specially Included Study*

2     We decided to include the results of an earlier investigation conducted by [8], where the box-counting fractal dimension was measured for RNF and FFA
3  images collected from patients at the University of Missouri School of Optometry. The results are summarized in **Table 1**.

4  **Table 1.** Outcomes of fractal analysis on patients as reported by *[8]*

| Pathology | Number of subjects | Mean (SD) fractal dimension for RNF images | Mean (SD) fractal dimension for FFA images | Significant difference in FD from normals? |
|---|---|---|---|---|
| Normal | 29 | 1.669 (0.0309) | 1.586 (0.0760) | N/A |
| Glaucoma | 25 | 1.694 (0.0229) | N/A | Yes |
| Ocular hypertension | 17 | 1.673 (0.1170) | N/A | No |
| Central Serous retinopathy | 11 | N/A | 1.592 (0.0352) | No |
| Diabetic retinopathy | 10 | N/A | 1.623 (0.0232) | Yes |
| Branch retinal vein occlusion | 4 | N/A | 1.597 (0.0258) | No |
| Central retinal vein occlusion | 3 | N/A | 1.569 (0.0470) | No |
| Central retinal artery occlusion | 2 | N/A | 1.557 (0.0254) | No |

5
6  *3.2 Demographic and Clinical Characteristics of Studies from Search*

7     9514 subjects were considered in total for this review. Studies focusing on diabetes mellitus mainly used random plasma glucose, Hb1Ac test, and/or diabetes
8  duration as diabetic factors when comparing fractal dimensions. All diabetic retinopathy studies differentiated between different stages of disease with the most
9  (5) using the ETDRS classification [21,22], 2 using the International Diabetic Retinopathy Severity Scale [23], and the remaining 3 not specifying a grading system.
10 Classification of myopia varied among the myopia studies. Some studies considered myopia to be defined as a spherical equivalent refraction (SER) worse than -6
11 diopters, while others set the boundary to be at -1.00 diopters. SER values between -1.00 and 1.00 diopters were generally considered emmetropic. Of the studies
12 which examined glaucoma, two used the International Society of Geographical and Epidemiological Ophthalmology scheme [24], while one used the Glaucoma
13 Hemifield Test [25], pattern standard deviation, and optic nerve damage as indicators but did not say whether it was part of some standardized procedure. Studies
14 on hypertension generally defined it as systolic blood pressure above 140 mmHg and diastolic blood pressure 80 mmHg. The clinical characteristics for each study
15 can be seen in **Table 2.** Most of the studies adjusted for potential confounding factors such as age, gender, cardiovascular risk factors, and optical disorders using
16 exclusion and/or multivariate regression. Age ranges varied between studies from schoolchildren as young at ages 11-12 to elderly adults as old as 71.

Table 2. Demographic information and clinical characteristics of study populations

| Author (year) | Study design (LOE) | Pathologies | N subjects Cases / controls | Mean (SD) Age in years: | Sex (M, F): Cases / controls | Population and clinical characteristics |
|---|---|---|---|---|---|---|
| Al-Sheikh et al. (2017) | Case-control | Myopia | 48<br>28/20 | Cases: 57.00 (17.93)<br>Controls: 56.05 (19.27) | (13 M, 15 F) /<br>(6 M, 14 F) | Myopic subjects with refraction greater than -6 diopters or axial lengths longer than 26.5 mm were included along with age-matched controls. Clinic where examinations took place was not stated. |
| Avakian et al. (2002) | Case-control | Diabetic Retinopathy | 9<br>5/4 | Not stated | Not stated | DR and control patients were diagnosed at the Ophthalmology Clinic, University of Washington, yielding 5 normal and 5 NPDR eyes. DR patients were diagnosed as having mild to moderate NPDR. |
| Azemin et al. (2014) | Case-control | Myopia | 82<br>41 / 41 | Cases: 23.93 (2.13)<br>Controls: 24.27 (2.49) | 35 M, 47 F total | Retinal images were collected from attendees at IIUM Optometry Clinic between January 2009 till June 2012. Classification into emmetropic and myopic groups was based on SER: emmetropia; between -1.00 D and +1.00 D SER, and myopia; SER of -3.00 D or worse. Subjects with lens opacity/history of ocular trauma/significant systematic disorder were excluded. |
| Bhardwaj et al. (2018) | Case-control | Diabetic Retinopathy | 61<br>35 / 26 | Cases: 56.3 (11.6)<br>Controls: 44.3 (19.1) | Not stated | Cases and controls were recruited from The New York Eye and Ear Infirmary of Mount Sinai, yielding 49 control eyes and 58 DR eyes. DR was graded according to the modified Arlie House classification of DR. Of the 58 DR eyes, 31 were categorized as NPDR (13 mild, 9 moderate, and 9 severe) and 27 as PDR. |
| Cheung et al. (2009) | Cross-sectional | Type 1 Diabetes Mellitus, Diabetic Retinopathy | 729<br>137 / 592 | Median age (IQR) of Total: 13.6 (12.8 – 15.0) | 322 M, 407 F total | Cases and controls were recruited from the Children's Hospital at Westmead, Sydney, Australia. DR was graded from ETDRS seven-field stereoscopic retinal photographs according to ETDRS adaptation of the Arlie House classification. Retinopathy was defined as ETDRS level 21 (minimal NPDR) or greater. |

| Author (year) | Study design (LOE) | Pathologies | N subjects Cases / controls | Mean (SD) Age in years: | Sex (M, F): Cases / controls | Population and clinical characteristics |
| --- | --- | --- | --- | --- | --- | --- |
| Cheung et al. (2012) | Cross-sectional | Diabetes, Diabetic Retinopathy, Hypertension, AMD, Myopia, Glaucoma | 2913 | Total: 57.67 (10.68) | 1393 M, 1520 F total | Data was taken from the baseline Singapore Malay Eye Study (SiMES-1) of urban Malay adults between 40 to 80 years of age. DM was defined as random plasma glucose ≥11.1 mmol/L, use of diabetic medication, or diagnosis from a physician. Hypertension was defined as blood pressure of 140/90 mmHg or more. Anterior chamber depth, spherical equivalent, and axial length were used as myopic measures. DR was graded using the ETDRS adaptation of Arlie House classification system, presence was defined as severity level of 15 or more. AMD was graded according to the Wisconsin Age-related Maculopathy Grading System and graded as present if early or late AMD signs were present. Glaucoma was diagnosed and classified using International Society of Geographical and Epidemiological Ophthalmology scheme. |
| Cheung et al. (2017) | Cohort | Diabetic Retinopathy | 427 | Total: 58.55 (8.66) | 186 M, 241 F total | Participants were eligible individuals who participated in the SiMES-1 baseline study and also the six year follow-up study SiMES-2. DR was graded as in Cheung et al. (2012). |
| Chicquet et al. (2020) | Case-control | Primary Open Angle Glaucoma | 122 61 / 61 | Median age (IQR) of 13.6 (1.8 – 15.0) for total population | (58 M , 64 F) / (50 M, 72 F) | Cases and controls were recruited from the University Hospital of Dijon and the University Hospital of Grenoble. Diagnosis was based on the Glaucoma Hemifield Test, pattern standard deviation, and characteristic optic nerve damage. |

| Author (year) | Study design (LOE) | Pathologies | N subjects Cases / controls | Mean (SD) Age in years: | Sex (M, F): Cases / controls | Population and clinical characteristics |
|---|---|---|---|---|---|---|
| Grauslund et al. (2010) | Cross-sectional | Type 1 Diabetes Mellitus, Diabetic Retinopathy | 178 | Total: 57.8 (12.6) | 111 M, 67 F total | Study was part of a cohort study of type 1 diabetes patients from Fye county, Denmark. Participants were identified based on insulin prescriptions in Fyn County, Denmark as described previously. Retinal photos were graded for according to ETDRS version of the Airlie House classification. PDR was considered present for levels 60 to 85. |
| Hirano et al. (2019) | Case-control | Diabetic Retinopathy | 62<br>46 / 16 | No DR: 52 (21)<br>NPDR: 62 (14)<br>PDR: 53 (14)<br>Controls: 47 (18) | (29 M, 17 F) /<br>(8 M, 8 F) | All participants were recruited from the Doheny Eye Center at the University of California, Los Angeles and the Shinshu University School of Medicine. DR severity was assessed using the International Clinical Diabetic Retinopathy Severity Scale. All cases were diabetic: 10 having no DR, 18 having NPDR, and 18 having PDR. |
| Kostic et al. (2018) | Case-control | Type 2 Diabetes Mellitus, Diabetic Retinopathy | 47<br>35 / 12 | No DR: 52.64 (7.79)<br>With DR: 52.63 (5.79)<br>Controls: 54.08 (7.71) | (14 M, 21 F) /<br>(3 M, 9 F) | Participants consisted of patients with DM that had DR up to ETDRS level 35 without macular edema, DM patients with no retinopathy, and healthy individuals. |
| Koulisis et al. (2016) | Case series | Retinal Venous Occlusion | 60<br>34 / 26 | Cases: 64.8 (8.8)<br>Controls: 60.9 (9.6) | (14 M, 20 F) /<br>(11 M, 15 F) | Study participants derived were adult subjects who underwent clinical evaluation and treatment for RVO. Control subjects were those without history of vascular disease or had it well controlled. Clinic/RVO diagnosis information not given. |
| Kunicki et al. (2009) | Case-control | Diabetic Retinopathy | 33<br>5 / 28 | Age range: 31-86 | Not stated | Control group consisted of 28 diabetic but ophthalmically normal patients. Case group consisted of 5 individuals with mild NPDR. All retinal images were obtained from the Digital Retinal Images for Vessel Extraction (DRIVE) database. |

| Author (year) | Study design (LOE) | Pathologies | N subjects Cases / controls | Mean (SD) Age in years: | Sex (M, F): Cases / controls | Population and clinical characteristics |
|---|---|---|---|---|---|---|
| Kurniawan et al. (2012) | Cohort | Hypertension | 1174 | Total: 11.91 (1.01) | 576 M, 598 F total | Subjects were schoolchildren from the Singapore Cohort Study of Risk Factors for Myopia. The systolic/diastolic blood pressures used for analysis was calculated by taking the average of three separate measurements. Mean arterial blood pressure was two thirds of diastolic blood pressure plus one third of systolic blood pressure. |
| Li et al. (2010) | Cross-sectional | Myopia | 2859 | Total: 64.17 (8.80) | 1247 M, 1612 F total | Study population was derived from Blue Mountains Eye Study. Emmetropia was defined as a SER between -1.00 D and 1.00 D. Myopia was classified as low (-1.01 D to -2.99 D), moderate (-3.00 D to -5.99 D), and high (-6.00 D or worse). |
| Li et al. (2017) | Case-control | Myopia | 40 20 / 20 | Cases: 28 (5) Controls: 30 (6) | (5 M, 15 F) / (7 M, 13 F) | Cases had SER worse than -5 D and controls had SER better than -3 D. Clinic where subjects were recruited from was not stated. |
| Lim et al. (2009) | Cohort | Diabetic Retinopathy | 590 | Age range: 12-20 | Not stated | Participants recruited from the Children's Hospital at Westmead, Sydney, Australia. Type 1 diabetes was defined following the Australasian Pediatric Endocrine Group diabetes register and national guidelines. Diabetic retinopathy was graded according to ETDRS adaptation of Arlie House classification. |
| Lim et al. (2017) | Cohort | Diabetic Retinopathy | 249 | Total: 59.9 (8.9) | 183 M, 66 F total | Subjects were adults with diabetes mellitus referred by physicians to Singapore National Eye Centre. Diabetic retinopathy was graded according to the ETDRS severity scale. |
| Olsen et al. (2015) | Cross-sectional | Type 2 Diabetes Mellitus | 103 | Total: 62.3 Range: 47.9-70.3 | 48 M, 55 F | Type 2 diabetic subjects were recruited from a DR screening clinic and examined at Odense University Hospital, Odense, Denmark. Only patients with no or minimal DR were invited. |

| Author (year) | Study design (LOE) | Pathologies | N subjects Cases / controls | Mean (SD) Age in years: | Sex (M, F): Cases / controls | Population and clinical characteristics |
|---|---|---|---|---|---|---|
| Remond et al. (2019) | Case series | NAION | 114<br>57 / 57 | Cases: 70 (8)<br>Controls: 71 (8) | (32 M, 25 F) /<br>(32 M, 25 F) | Study included retrospectively imaged patients between September 2007 and July 2017 at Grenoble Alpes University. |
| Tang et al. (2017) | Cross-sectional | Type 1 or 2 Diabetes Mellitus,<br>Diabetic Retinopathy | 286 | Total: 65.69 (10.73) | 122 M, 164 F total | DM (type 1 or 2) patients at CUHK Eye Centre, Hong Kong Eye Hospital |
| Tang et al. (2020) | Cross-sectional | Type 1 or 2 Diabetes Mellitus,<br>Diabetic Retinopathy | 250 | Total: 60.7 (13.2) | 135 M, 115 F total | DM (type 1 or 2) patients at CUHK Eye Centre, Hong Kong Eye Hospital |
| Ting et al. (2017) | Cohort | Diabetic Retinopathy | 50 | Total: 59.5 (8.9) | 26 M, 24 F total | Type 2 diabetic subjects recruited from DR screening clinic in the Singapore National Eye Center, Singapore, Singapore. DR was graded according to the International Clinical Diabetic Retinopathy Severity Scales. |
| Tzaridis et al. (2019) | Cross-sectional | MacTel Type 2 | 100<br>76 / 24 | Cases: 62.3 (6.1)<br>Controls: 61.5 (5.9) | Not stated | Subjects with confirmed diagnosis were recruited from the Natural History and Observation Study (NHOS) in Germany at the Department of Ophthalmology, University Hospital of Bonn, Germany. |
| Wu et al. (2013) | Cross-sectional | Glaucoma | 2789<br>123 / 2666 | Age range: 40-80 | Not stated | Subjects derived from SiMES-1. Glaucoma was diagnosed according to the International Society of Geographic and Epidemiological Ophthalmology criteria. Ocular hypertension (OHT) was defined as intraocular pressure > 21.5 mmHg without glaucoma. |

| Author (year) | Study design (LOE) | Pathologies | N subjects Cases / controls | Mean (SD) Age in years: | Sex (M, F): Cases / controls | Population and clinical characteristics |
|---|---|---|---|---|---|---|
| Xu et al. (2020) | Case-control | Hypertension | 120<br>77 / 43 | Cases: 59.2 (7.6)<br>Controls: 57.0 (6.8) | (26 M, 51 F) /<br>(15 M, 28 F) | Subgroup of an observational study population that was conducted in the village area of Pingyin County in the southwest of Jinan, Shandong, China. Inclusion in the hypertensive group was determined by having blood pressure > 140/90 mmHg and < 180/110 mmHg, age from 40-80 years, and no signs of confounding comorbidities. Mean arterial pressure defined as diastolic blood pressure plus one third of the diastolic and systolic blood pressure difference. |
| Yang et al. (2016) | Case-control | Myopia | 45<br>21 / 24 | Cases: 26.0 (2.7)<br>Controls: 27.4 (6.4) | (14 M, 10 F) /<br>(12 M, 9 F) | Patients from Wenzhou Eye Hospital and students from Wenzhou Medical University were recruited. Control group consisted of subjects with SER between +0.50 D and -3.00 D and highly myopic group consisted of subjects with SER less than -6.00 D. |
| Zahid et al. (2016) | Case series | Diabetic Retinopathy | 29<br>8 / 21 | Cases: 56.66 (12.65)<br>Controls: 32.27 (6.71) | Not stated | Subjects were recruited from the private practices of Vitreous, Retina, Macula Consultants of New York. Case group inclusion criteria included age greater than 18 years, history of DM, and presence of mild NPDR to PDR without evidence of DME. Control group included volunteer subjects without history of diabetes or ocular disease. |

*3.3 Results of Studies Conducting Fractal Analysis on Retinal Images*

Changes in the FD of the retinal vasculature in patients with retinal disorders are reported in **Table 3**. Setups for retinal imaging, vessel segmentation, and FD calculation were varied. Methodological variance was especially prominent for studies that employed OCTA as the imaging method, as often different studies considered different retinal vascular layers when calculating FD. Some studies also calculated FD on only the veins or arteries in an image, leading to the distinct parameters venular and arteriolar FD, apart from total FD. Digital fundus photography was the most common retinal imaging method, which was used by 14 out of the 28 studies with the most common field of view (FOV) being 45 degrees. The next most common imaging method was OCTA, which was used by 13 studies. Only one study used fundus fluorescein angiography (FFA) as the imaging method. Although methodological setups were relatively varied across the studies, there were some similarities. One setup that was used by 5 of the studies was the use of the Singapore I Vessel Assessment (SIVA, National University of Singapore, Singapore) software combined with an optic disc centered annular ROI of 0.5 to 2.0 disc diameters. Another was the use of digital retinal photography combined with the fractal analysis package of the International Retinal Imaging Software (IRIS) for a circular ROI of 3.5.disc diameters centered at the optic disc, which was the choice of 4 other studies. Most of the studies (25 out of 28) calculated just the monofractal box-counting dimension of the retinal vascular tree and did not consider its multifractal properties. It should also be noted that 22 of the 28 studies examined retinal vascular parameters other than fractal dimension such as branching angles, caliber, and tortuosity. The study by [26] was unique among the rest in that it examined the relationship between FD and a variety of different pathologies for the Singapore Malay Eye Study (SiMES) cohort. [26] found a significant, independent association between decreased retinal FD and morbidities such as blood pressure and myopia. Glaucoma and Age-related macular degeneration were not significantly associated with FD and an association between diabetic retinopathy and FD was ruled out after multivariate regression analysis [26]. For the studies that looked at myopia [26–31], hypertension [26,32,33], glaucoma, and diabetes [34–38], all generally observed a mean decrease in fractal dimension relative to the control group or a decreasing trend with respect to increasing disease severity. In contrast, findings from the diabetic retinopathy studies were more mixed. Decreased retinal FD was found to be associated with presence of NPDR [39,40] and PDR [40]. Retinal FD also showed a decreasing trend with respect to diabetic retinopathy severity ranging from mild NPDR to PDR [37,38,41,42]. However, three studies found that greater retinal FD was associated with early retinopathy signs in young type 1 diabetic patients [43], incidence of referable diabetic retinopathy in urban Malay adults [44], and presence of proliferative diabetic retinopathy in type 2 diabetes [45]. For retinographies from the DRIVE database, no significant change in FD was reported for patients with mild NPDR for the entire retina although a significant difference was found for the macular region [46]. A cohort study [47] also reported no association found between retinal FD and incident diabetic retinopathy in young diabetes patients after a mean (SD) follow-up period of 2.9 (2.0) years. For the studies that examined less common retinal diseases, [48] found decreased FD for OCTA images of the superficial and deep retinal layers in patients with retinal occlusions, [49] found an increase in arteriolar/venular FDs in NAION patients, and [50] found FD decreases in the deep/superficial retinal plexuses in MacTel patients.

**Table 3.** Outcomes of fractal analysis for retinal pathology

| Author (year) | RVPs | Imaging device | FD ROI | FD Calculation | Results |
| --- | --- | --- | --- | --- | --- |
| Al-Sheikh et al. (2017) | FDt, vessel density | Optovue RTVue-XR Avanti; OptoVue, Inc., Fremont CA, USA | Whole area permitted by 3x3 mm macula scan | Box-counting method using Fractalyse (ThéMA, Besançon Cedex, France). | Myopic eyes had statistically significantly lower FD in both SCP and DCP layers. |
| Avakian et al. (2002) | FDt, grid intersection | 60° fundus fluorescein angiography | RVPs were calculated for the macular and paramacular regions and also for the entire retina. The image was divided into nine square, equal zones with the central square denoted the macular region, and the paramacular region being comprised of the five zones above, below, and to the right of the macular region. | Segmentation was done manually using NIH, Adobe Photoshop, and computerized processing. RVPs calculation was done using a custom-written computer program. | FD was found to be significantly higher in the macular region for normal patients as opposed to NPDR patients. FD difference in paramacular regions were not strongly significant. |
| Azemin et al. (2014) | FDt | 45° digital retinal camera (AFC-230/210) | Whole area of the image, which is centered at the midway point between the macula and OD. | Vessel segmentation was done using custom-written software. Box-counting FD was calculated using the FracLac plugin of ImageJ. | Eyes with myopia tended to have a smaller FD than the emmetropic control group. |
| Bhardwaj et al. (2018) | FD_SCP, FD_DCP, correlation coefficient | Optovue RTVue-XR Avanti; OptoVue, Inc., Fremont CA, USA | Whole area permitted by 3x3 mm scan centered about the fovea. | Box-counting method using Fractalyse. | For all stages of DR, FD was found to be reduced in diabetic eyes as opposed to controls. The DCP was more consistently affected than the SCP. |
| Cheung et al. (2009) | FDt | Seven-standard field stereoscopic retinal photography with Topcon Fundus Camera (TRC 50-VT; Tokyo Optical, Tokyo, Japan) | Circular region of 3.5 DD centered on the OD. | Vessel segmentation and box-counting FD with IRIS-Fractal | Increased FD was associated with early retinopathy signs in young individuals. |

| Author (year) | RVPs | Imaging device | FD ROI | FD Calculation | Results |
| --- | --- | --- | --- | --- | --- |
| Cheung et al. (2012) | FDt, CRAE, CRVE | 45° digital retinal camera (Canon CR-DGi with a 10D SLR digital camera backing; Canon, Tochigiken, Japan) | 0.5-2.0 disc diameters away from the OD margin. | RVPs were calculated using Singapore I Vessel Assessment (SIVA) semi-automatic computer program. FD was calculated using box-counting. | Smaller FD was associated with older age, diabetes, hypertension, myopia refraction, and presence of cataract. AMD, glaucoma, had no impact on FD. DR was not related to FD after performing multivariate regression analysis. |
| Cheung et al. (2017) | FDa/v, BAa/v, BCa/v, CRAE, CRVE, sTORTa/v, cTORTa/v | 45° digital retinal camera (Canon CR-DGi 10D; Canon, Tokyo, Japan) | 0.5-2.0 disc diameters away from the OD margin. | Box-counting using SIVA | Higher values of arteriolar fractal dimension were associated with incidence of referable DR. |
| Chicquet et al. (2020) | CRAE, CRVE, TORTa/b, AVR, FDa/v/t (multifractal: capacity, information, correlation) | 30 or 40° digital retinal camera (Visu-cam 200; Carl Zeiss Meditec, Oberkochen, Germany) or (CR2; Canon, Europa, Amstelveen, The Netherlands) | Centered at the OD, circular regions: Zone A [0-0.5 OD disc diameters], Zone B [0.5-1 disc diameters], Zone C [0.5-2.0 disc diameters]. | Fractal parameters were calculated in Zone C using VAMPIRE software. | POAG patients exhibited reduced values of fractal parameters. |
| Grauslund et al. (2010) | FDt | 45° digital retinal camera (TRC-NW6S; Topcon, Tokyo, Japan) | Circular region of 3.5 disc diameters centered at the OD. | Box-counting method using IRIS-Fractal | Persons with lower FD were more likely to have proliferative retinopathy, neuropathy, and nephropathy. |
| Hirano et al. (2019) | Perfusion density, Vessel length density, FDt | PLEX Elite 9000; Carl Zeiss Meditec, Dublin, California, USA | 3x3, 6x6, and 12x12 mm fovea centered scans. Unsegmented retina, SRL, and DRL layers considered separately. | Box-counting method using Fractalyse. | FD progressively decreased with worsening DR severity in segmented and non-segmented layers for all scan sizes. |
| Kostic et al. (2018) | FDt, FAZ perimeter, FAZ roundness, FAZ area | 20° digital retinal camera, exact model not stated | Whole area permitted by photo. | Box-counting method was used after importing images into ImageJ and grayscale format conversion | The healthy group had the highest FD compared to the DM (Diabetes Mellitus) and MDR (Mild diabetic retinopathy) groups. FD in the MDR group was also significantly lower than the DM group. |

| Author (year) | RVPs | Imaging device | FD ROI | FD Calculation | Results |
|---|---|---|---|---|---|
| Koulisis et al. (2016) | Skeletal density, vessel diameter index, vessel density, FDt, FD_SRL, FD_DRL | Cirrus; Carl Zeiss Meditec, Dublin, CA, USA | 3x3 mm scan centered on the fovea. Unsegmented retina, SRL, and DRL layers considered separately. | Box-counting method using custom software. | Lower mean FD was demonstrated in the non-segmented layer of RVO eyes compared to controls. Subgroup analysis of CRVO and BRVO eyes independently demonstrated lower mean non-segmented FD versus controls. FD was significantly lower in CRVO, BRVO eyes compared to the unaffected fellow eye. In SRL, CRVO and BRVO eyes demonstrated significantly lower FD. In DRL, only CRVO eyes had this trend. |
| Kunicki et al. (2009) | FD0 (Box-counting), FD1 (Information) | 45° digital retinal camera (CR5; Canon, Tokyo, Japan) | Image was divided into 9 square regions of equal size. Fractal dimensions were calculated in each of those regions as well for the entire retina. | Fractal parameters calculated using Benoit Fractal Analysis System (TruSoft, USA) | Results did not show significant difference between cases and controls. |
| Kurniawan et al. (2012) | CRAE, CRVE, FD | Digital retinal photography, model and FOV not stated. | Circular region of 3.5 disc diameters centered at the OD. | Box-counting using IRIS-Fractal | Blood pressure was inversely related to retinal FD. |
| Li et al. (2010) | FDt | 30° digital retinal camera (Zeiss FF3; Carl Zeiss, Oberkochen, Germany) | Circular region of 3.5 disc diameters centered at the OD. | Box-counting using IRIS-Fractal | Increasing myopia severity was associated with reduced FD after adjusting for confounders, which might suggest rarefaction of retinal vasculature associated with high myopia. |

| Author (year) | RVPs | Imaging device | FD ROI | FD Calculation | Results |
| --- | --- | --- | --- | --- | --- |
| Li et al. (2017) | FD_SVP, FD_DVP | Zeiss HD-OCT with Angioplex; Carl Zeiss Meditec, Oberkochen, Germany | Annular zone 0.6-2.5 mm in diameter. The annular zone was further divided into 4 quadrantal zones called the superior temporal, inferior temporal, superior nasal, and inferior nasal. The annular zone was also divided into 6 thin annuli with width ~0.16 mm denoted C1-C6. | Box-counting using Benoit Pro 2.0 (TruSoft International Inc, St Petersburg, FL) | The FD of the superficial and deep microvascular networks were significantly decreased in the myopia group in comparison to the controls. |
| Lim et al. (2009) | FDt | Digital retinal camera (TRC 50-VT; Topcon, Tokyo, Japan), FOV not stated | Circular region of 3.5 disc diameters centered at the OD. | Box-counting method using IRIS-Fractal | Retinal vascular FD was not associated with incident early DR in the sample of children/adolescents with type 1 diabetes. |
| Lim et al. (2017) | FDa/v, cTORTa/v, BCa/v, BAa/v | 45° digital retinal camera (Canon CR-DGi with a 10D SLR digital camera backing; Canon, Tochigiken, Japan) | Circular region of 0.5-2.0 disc diameters from the disc margin. | Box-counting using SIVA | Larger venular FDs were associated with increased likelihood of DR incidence. |
| Olsen et al. (2015) | FDt | Optos 200Tx (Optos plc, Dunferm-line, Scotland, UK) | Circular region of 0.5-2.0 disc diameters from disc margin. | Box-counting using SIVA | Correlation found between lower FD and presence of diabetic neuropathy. |
| Remond et al. (2019) | AVR, CRVE, CRAE, TORTa/v, FDa/v (multifractal: capacity, information, correlation) | 45° digital retinal camera (TRC 50 IX; Topcon, Tokyo, Japan) | Circular region of 0.5-2.0 disc diameters from disc margin. | Fractal parameters calculated using VAMPIRE | Fractal dimension D0a was significantly higher in the NAION group compared to the control group but did not change significantly at different disease stages. |
| Tang et al. (2017) | FAZ area, vessel density, FD_DCP | Triton DRI-OCT; Topcon, Inc., Tokyo, Japan | Whole area permitted by 3x3 mm scan centered on the fovea. | Box-counting using customized MATLAB program | DR severity had the most impact on the RVPs and was found to be associated with lower FD. |

| Author (year) | RVPs | Imaging device | FD ROI | FD Calculation | Results |
| --- | --- | --- | --- | --- | --- |
| Tang et al. (2020) | FAZ area, FAZ circularity, total vessel density, parafoveal vessel density, FD_SCP, vessel diameter index | Triton DRI-OCT; Topcon, Inc., Tokyo, Japan | Whole area permitted by 3x3 mm scan centered on the fovea. | Box-counting using customized MATLAB program | DR severity was found to be associated with lower FD. |
| Ting et al. (2017) | Capillary density index, FD_SVP, FD_DVP | Swept-source OCT (Topcon Corp) | Whole area permitted by 3x3 mm scan centered on the fovea. | Box-counting after binarization with ImageJ | FD was positively correlated with worsening DR severity. |
| Tzaridis et al. (2019) | Vessel density, skeleton density, FD_SVP, FD_DVP | PLEX Elite 9000; Carl Zeiss Meditec, Dublin, CA, USA | Annular region was defined between 0.9 and 2.4 diameter circles centered on the fovea. The annular region was then divided into four quadrants by two diagonal perpendicular lines. Fractal analysis were conducted in the rightmost quadrant called the temporal parafovea. | Box-counting with FracLac | FD in the temporal parafovea of the deep retinal plexus showed a progressive decrease in all disease stages. A significant decrease in FD was also found for the deep plexus nasal sector for advanced and neovascular stages. In the superficial plexus, FD was significantly decreased in the temporal parafovea in the advanced and neovascular stages but did not differ from controls in the early stages. |
| Wu et al. (2013) | BAa/v, TORTa/v, FDt | 45° digital retinal camera (Canon CR-DGi with a 10D SLR digital camera backing; Canon, Tokyo, Japan) | Circular region of 0.5-2.0 disc diameters from the disc margin. | Box-counting using SIVA | Eyes with glaucoma had significantly lower FD. Similarly, eyes with ocular hypertension had significantly lower FD. FD was also lower in eyes with higher IOP. |

| Author (year) | RVPs | Imaging device | FD ROI | FD Calculation | Results |
| --- | --- | --- | --- | --- | --- |
| Xu et al. (2020) | Vessel density, skeleton density, vessel diameter index, FD, FAZ area, retinal vessel caliber (calculated for both superficial and deep retinal layers) | Optovue RTVue XR Avanti; Optovue Inc., Fremont, CA, USA | Whole area permitted by 6x6 mm scan centered on the fovea. | Box-counting method with Fractalyse | FD of the SRL and DRL showed significant reduction when comparing hypertensive eyes with controls. |
| Yang et al. (2016) | FDt, FD_SVP, FD_DVP. All repeated with separation of large from micro vessels | Optovue RTVue XR Avanti; Optovue Inc., Fremont, CA, USA | Annular zone of 0.6-2.5 mm in diameter. The annular zone was further divided into four quadrantal zones called the superior temporal, inferior temporal, superior nasal, and inferior nasal. The annular zone was also divided into 6 thin annuli with width ~0.16 mm denoted C1-C6. | Box-counting method using Benoit fractal analysis toolbox (TruSoft Internation, Inc., St. Petersburg, FL, USA) | Densities in six annular zones and four quadrantal zones of the superficial, deep, and whole retinal layers were significantly lower in highly myopic subjects compared to controls. |
| Zahid et al. (2016) | FD_SCP, FD_DCP, vessel density | Optovue RTVue XR Avanti; Optovue Inc., Fremont, CA, USA | Whole area permitted by 3x3 mm scan centered on the fovea. | Box-counting method with Fractalyse | The average FD for diabetic eyes was significantly lower than control eyes for the superficial and deep capillary plexuses. |

**Table 4.** Results of meta-analysis for retinal disease subgroups

| Disease | Authors (year) | No. of cases | No. of normals | Mean (SD) FD - Cases | Mean (SD) FD - Normals | Effect Size (SE) | Summary Effect Size (SE) and 95% C.I. |
|---|---|---|---|---|---|---|---|
| DR | Avakian et al. (2002) | 5 | 4 | 1.41 (0.02) | 1.46 (0.02) | -0.05 (0.0134) | -0.00267 (0.0088) [-0.0439, -0.0095] |
|  | Bhardwaj et al. (2018) | 35 | 26 | 1.581 (0.0763) | 1.664 (0.0546) | -0.083 (0.0161) |  |
|  | Cheung et al. (2009) | 137 | 592 | 1.4676 (0.01) | 1.4598 (0.02) | 0.0078 (0.0018) |  |
|  | Cheung et al. (2012) | 262 | 2648 | 1.4054 (0.0412) | 1.4053 (0.0421) | 0.0001 (0.0027) |  |
|  | Hirano et al. (2019) | 36 | 16 | 1.482 (0.0341) | 1.534 (0.010) | -0.052 (0.0087) |  |
|  | Kostic et al. (2018) | 15 | 12 | 1.35 (0.03) | 1.42 (0.03) | -0.07 (0.0116) |  |
|  | Kunicki et al. (2009) | 5 | 28 | 1.462 (0.021) | 1.470 (0.025) | -0.008 (0.0119) |  |
|  | Lakshminarayanan et al. (2002) | 10 | 29 | 1.623 (0.0232) | 1.586 (0.0760) | 0.037 (0.0246) |  |
|  | Ting et al. (2017) | 41 | 9 | 1.5899 (0.06) | 1.54 (0.0546) | 0.0499 (0.0218) |  |
|  | Zahid et al. (2016) | 8 | 21 | 1.5815 (0.0565) | 1.6785 (0.0588) | -0.097 (0.0242) |  |
| DM | Cheung et al. (2012) | 667 | 2163 | 1.4063 (0.0439) | 1.402 (0.0419) | 0.0043 (0.0019) | -0.0118 (0.0171) [-0.0453, 0.0218] |
|  | Kostic et al. (2018) | 20 | 12 | 1.39 (0.02) | 1.42 (0.03) | -0.03 (0.0088) |  |
| Myopia | Al-Sheikh et al. (2017) | 28 | 20 | 1.587 (0.0385) | 1.609 (0.0209) | -0.022 (0.0095) | -0.0176 (0.0053) [-0.0279, -0.0073] |
|  | Azemin et al. (2014) | 41 | 41 | 1.5588 (0.0142) | 1.5666 (0.0160) | -0.0078 (0.0033) |  |
|  | Li et al. (2010) | 281 | 1380 | 1.437 (0.0255) | 1.444 (0.0189) | -0.007 (0.0013) |  |
|  | Li et al. (2017) | 20 | 20 | 1.705 (0.0603) | 1.75 (0.0156) | -0.045 (0.0139) |  |
|  | Yang et al. (2016) | 21 | 24 | 1.783 (0.014) | 1.809 (0.008) | -0.026 (0.0033) |  |
| Hypertension | Cheung et al. (2012) | 1949 | 963 | 1.404 (0.0441) | 1.4082 (0.0465) | -0.0042 (0.0018) | -0.0054 (0.0036) [-0.0125, 0.0017] |
|  | Xu et al. (2020) | 77 | 43 | 1.6895 (0.0689) | 1.706 (0.0346) | -0.0165 (0.0112) |  |
| Glaucoma | Cheung et al. (2012) | 72 | 2841 | 1.4067 (0.0433) | 1.4054 (0.0426) | 0.0013 (0.0051) | -0.0049 (0.0191) [-0.0423, 0.0324] |
|  | Lakshminarayanan et al. (2002) | 25 | 29 | 1.694 (0.0229) | 1.669 (0.0309) | 0.025 (0.0075) |  |
|  | Wu et al. (2013) | 123 | 2666 | 1.37 (0.10) | 1.41 (0.04) | -0.04 (0.0041) |  |

*3.4 Results of Meta-Analysis*

The results of the meta-analysis are reported in **Table 4**. Meta-analysis was conducted for the disease categories: diabetic retinopathy, diabetes mellitus, myopia, hypertension, and glaucoma. Studies examining retinal occlusions, AMD, NAION, and MacTel were excluded due to those categories each having less than two studies found from the search. Due to some studies calculating mean (SD) FD for different disease severities or retinal vascular layers, all groups of mean (SD) FD were combined into a single group. Study effect sizes were calculated by subtracting the normal group mean from the case group mean. Due to heterogeneity between studies, a random effect model [20] was used to synthesize all study effect sizes to derive a summary effect size and its 95% confidence interval for each disease category. The diabetic retinopathy and myopia categories had the most data available and had summary effect sizes [95% C.I.] of -0.00267 [-0.0439, -0.0095] and -0.0176 [-0.0279, -0.0073] respectively, suggesting decreased fractal dimension for those pathologies. The other categories: diabetes Mellitus, hypertension, and glaucoma had summary effect sizes [95% C.I.] of -0.0118 [-0.0453, 0.0218], -0.0054 [-0.0125, 0.0017], and -0.0049 [-0.0423, 0.0324] respectively with the relationship remaining uncertain due the confidence intervals encompassing positive effect sizes as well.

**4. Discussion**

*4.1 – Limitations/Stability of Fractal Dimension*

The outcomes of fractal analysis show decreased retinal FD in myopia, glaucoma, and hypertension but show inconsistent results for diabetic retinopathy. Other diseases that were considered in this review: AMD, MacTel, NAION, and retinal occlusions were not investigated extensively enough in the literature for meta-analysis or to form a conclusive consensus. It is difficult to compare fractal analysis outcomes due to heterogeneity in methods. Often studies considered different regions of interest and vascular layers, as well as ignoring certain types of blood vessels instead of calculating over the entire tree. Variation in imaging methods, image processing, and fractal analysis tools is also concerning when comparing research between studies. Previously, [12] studied the robustness of the FD parameter with respect to different methodological setups and concluded that retinal FD can be misleading in clinical applications due to its sensitivity to image quality and technique used. Similarly, [51] found a significant dependence of FD on the vessel segmentation and dimensional calculation methods used. Even lesser factors such as image brightness, contrast, and focus can have significant impact on the final FD estimate [52]. [53] conducted a review on the association between retinal FD and neurodegenerative diseases: Alzheimer's, cognitive impairment, and stroke. Although a general decrease in retinal FD for patients with neurodegenerative pathology was observed, [53] also expressed difficulties with study comparison and called for the uniformization and standardization of procedures related to calculating retinal FD before establishing clinical applications.

*4.2 – Meta-analysis*

To answer the question of whether retinal FD changes significantly in a quantitative manner, a meta-analysis was conducted for each retinal disease category to derive a summary effect size using the available studies. A negative summary effect size was found for diabetic retinopathy and myopia, suggesting decreased retinal fractal dimension for patients with those pathologies while the results for diabetes mellitus, glaucoma, hypertension are uncertain. Some limitations of this analysis include combining the retinal vascular layers into one group, estimation of mean (SD) FD, study heterogeneity, as well as low sample size for diabetes, glaucoma, and hypertension. Taking the combined average of mean (SD) FD calculations over different retinal vascular layers is a source of error as the fractal dimension over the whole retina can be greater than over its constituent layers. Estimating the mean (SD) FD for a study given median (IQR) FD can also lead to errors in the meta-analysis. Furthermore, the demographics of each study vary widely in age and ocular history. This combined with the many variables that go into each stage of data acquisition/processing when performing fractal analysis leads to large heterogeneity. Because of this, we reiterate that a standardized procedure for retinal fractal analysis should be developed to facilitate inter-study comparison and support future meta-analyses.

5. Conclusion

This review summarizes the current scientific literature on the association between FD and retinal disease. The nature of the association depends on the type of retinal disease in consideration. The results of the qualitative synthesis show decreased fractal dimension associated with presence of glaucoma, hypertension, and myopia. However, the results of the meta-analysis show that the decrease is strongest with diabetic retinopathy and myopia and weak for diabetes, glaucoma, hypertension. Due to the variances in methodological setups for retinal image processing and FD calculation, it is difficult to form a consensus on effect. Hence, before moving onto clinical applications of FD, it is necessary that a standardized protocol for image acquisition/processing be established first to facilitate inter-study comparison.

**Abbreviations:**

The following abbreviations were used in this manuscript:
AMD – age-related macular degeneration
AVR – arteriolar to venular diameter ratio
BAa/v - arteriolar/venular branching angle
BCa/v – arteriolar/venular branching coefficient
CRAE – central retinal arteriolar equivalent
CRVE – central retinal venular equivalent
cTORTa/v – curvature arteriolar/venular tortuosity
DCP – deep capillary plexus
DM – diabetes mellitus
DR – diabetic retinopathy
DRL – deep retinal layer
DVP – deep vascular plexus, FD – fractal dimension
FAZ – foveal avascular zone
FFA – fundus fluorescein angiography
FOV – field of view
FDa/v/t – arteriolar/venular/total fractal dimension
IOP – intraocular pressure
MacTel – macular telangiectasia
NAION – nonarteritic anterior ischemic optic neuropathy
NPDR – non-proliferative diabetic retinopathy
OCTA – optical coherence tomography angiography
OD – optic disc
PDR – proliferative diabetic retinopathy
POAG – primary open angle glaucoma
RNF – retinal nerve fiber
ROI – region of interest
RVP – retinal vessel parameter
SCP – superficial capillary plexus
SD – standard deviation
SRL – superficial retinal layer
SVP – superficial vascular plexus
TORTa/v – arteriolar/venular simple tortuosity

**Funding:** S.Y. received funding from the Government of Canada's Student Work Placement Program (SWPP) to do a four-month co-op with the Faculty of Science, University of Waterloo under the supervision of V.L., V.L. also acknowledges a Discovery grant from The Natural Sciences and Engineering Research Council of Canada.

**Conflicts of Interest:** The authors declare no conflict of interest.

**References**

1. Murray, C.D. The Physiological Principle of Minimum Work: I. The Vascular System and the Cost of Blood Volume. *Proc. Natl. Acad. Sci.* **1926**, *12*, 207–214, doi:10.1073/pnas.12.3.207.
2. Daxer, A. Characterisation of the Neovascularisation Process in Diabetic Retinopathy by Means of Fractal Geometry: Diagnostic Implications. *Graefes Arch. Clin. Exp. Ophthalmol. Albrecht Von Graefes Arch. Klin. Exp. Ophthalmol.* **1993**, *231*, 681–686, doi:10.1007/BF00919281.
3. Daxer, A. The Fractal Geometry of Proliferative Diabetic Retinopathy: Implications for the Diagnosis and the Process of Retinal Vasculogenesis. *Curr. Eye Res.* **1993**, *12*, 1103–1109.
4. Family, F.; Masters, B.R.; Platt, D.E. Fractal Pattern Formation in Human Retinal Vessels. *Phys. Nonlinear Phenom.* **1989**, *38*, 98–103, doi:10.1016/0167-2789(89)90178-4.
5. Mainster, M.A. The Fractal Properties of Retinal Vessels: Embryological and Clinical Implications. *Eye Lond. Engl.* **1990**, *4 ( Pt 1)*, 235–241, doi:10.1038/eye.1990.33.
6. Campbell, J.P.; Zhang, M.; Hwang, T.S.; Bailey, S.T.; Wilson, D.J.; Jia, Y.; Huang, D. Detailed Vascular Anatomy of the Human Retina by Projection-Resolved Optical Coherence Tomography Angiography. *Sci. Rep.* **2017**, *7*, 42201, doi:10.1038/srep42201.
7. Mandelbrot, B.B.; Wheeler, J.A. The Fractal Geometry of Nature. *Am. J. Phys.* **1983**, *51*, 286–287, doi:10.1119/1.13295.
8. Lakshminarayanan, V.; Raghuram, A.; Myerson, J.; Varadharajan, S. The Fractal Dimension in Retinal Pathology. *J. Mod. Opt. - J MOD Opt.* **2003**, *50*, 1701–1703, doi:10.1080/09500340310000069442.
9. Soares, J.V.B.; Leandro, J.J.G.; Cesar, R.M.; Jelinek, H.F.; Cree, M.J. Retinal Vessel Segmentation Using the 2-D Gabor Wavelet and Supervised Classification. *IEEE Trans. Med. Imaging* **2006**, *25*, 1214–1222, doi:10.1109/TMI.2006.879967.
10. Zhang, J.; Bekkers, E.; Abbasi, S.; Dashtbozorg, B.; ter Haar Romeny, B. Robust and Fast Vessel Segmentation via Gaussian Derivatives in Orientation Scores. In Proceedings of the Image Analysis and Processing — ICIAP 2015; Murino, V., Puppo, E., Eds.; Springer International Publishing: Cham, 2015; pp. 537–547.
11. Staal, J.; Abramoff, M.D.; Niemeijer, M.; Viergever, M.A.; Ginneken, B. van Ridge-Based Vessel Segmentation in Color Images of the Retina. *IEEE Trans. Med. Imaging* **2004**, *23*, 501–509, doi:10.1109/TMI.2004.825627.
12. Huang, F.; Dashtbozorg, B.; Zhang, J.; Bekkers, E.; Abbasi-Sureshjani, S.; Berendschot, T.T.J.M.; Ter Haar Romeny, B.M. Reliability of Using Retinal Vascular Fractal Dimension as a Biomarker in the Diabetic Retinopathy Detection. *J. Ophthalmol.* **2016**, *2016*, 6259047, doi:10.1155/2016/6259047.
13. Rényi, A. On the Dimension and Entropy of Probability Distributions. *Acta Math. Acad. Sci. Hung.* **1959**, *10*, 193–215, doi:10.1007/BF02063299.
14. Tălu, S. Multifractal Geometry in Analysis and Processing of Digital Retinal Photographs for Early Diagnosis of Human Diabetic Macular Edema. *Curr. Eye Res.* **2013**, *38*, 781–792, doi:10.3109/02713683.2013.779722.

15. Ott, E. *Chaos in Dynamical Systems*; Cambridge University Press: Cambridge England; New York, NY, USA, 1993; ISBN 978-0-521-43799-8.
16. Moher, D.; Liberati, A.; Tetzlaff, J.; Altman, D.G.; Group, T.P. Preferred Reporting Items for Systematic Reviews and Meta-Analyses: The PRISMA Statement. *PLOS Med.* **2009**, *6*, e1000097, doi:10.1371/journal.pmed.1000097.
17. Schardt, C.; Adams, M.B.; Owens, T.; Keitz, S.; Fontelo, P. Utilization of the PICO Framework to Improve Searching PubMed for Clinical Questions. *BMC Med. Inform. Decis. Mak.* **2007**, *7*, 16, doi:10.1186/1472-6947-7-16.
18. Luo, D.; Wan, X.; Liu, J.; Tong, T. Optimally Estimating the Sample Mean from the Sample Size, Median, Mid-Range, and/or Mid-Quartile Range. *Stat. Methods Med. Res.* **2018**, *27*, 1785–1805, doi:10.1177/0962280216669183.
19. Wan, X.; Wang, W.; Liu, J.; Tong, T. Estimating the Sample Mean and Standard Deviation from the Sample Size, Median, Range and/or Interquartile Range. *BMC Med. Res. Methodol.* **2014**, *14*, 135, doi:10.1186/1471-2288-14-135.
20. Sutton, A.; Abrams, K.R.; Jones, D.R.; Sheldon, T.; Song, F. *Methods for Meta-Analysis in Medical Research*; Wiley, 2000; p. 317; ISBN 978-0-471-49066-1.
21. Diabetic Retinopathy Study. Report Number 6. Design, Methods, and Baseline Results. Report Number 7. A Modification of the Airlie House Classification of Diabetic Retinopathy. *Invest. Ophthalmol. Vis. Sci.* **1981**, *21*, 1–226.
22. Early Treatment Diabetic Retinopathy Study Research Group. Grading Diabetic Retinopathy from Stereoscopic Color Fundus Photographs--an Extension of the Modified Airlie House Classification. ETDRS Report Number 10. *Ophthalmology* **1991**, *98*, 786–806.
23. Wilkinson, C.P.; Ferris, F.L.; Klein, R.E.; Lee, P.P.; Agardh, C.D.; Davis, M.; Dills, D.; Kampik, A.; Pararajasegaram, R.; Verdaguer, J.T.; et al. Proposed International Clinical Diabetic Retinopathy and Diabetic Macular Edema Disease Severity Scales. *Ophthalmology* **2003**, *110*, 1677–1682, doi:10.1016/S0161-6420(03)00475-5.
24. Foster, P.J.; Buhrmann, R.; Quigley, H.A.; Johnson, G.J. The Definition and Classification of Glaucoma in Prevalence Surveys. *Br. J. Ophthalmol.* **2002**, *86*, 238–242.
25. Asman, P.; Heijl, A. Glaucoma Hemifield Test. Automated Visual Field Evaluation. *Arch. Ophthalmol. Chic. Ill 1960* **1992**, *110*, 812–819, doi:10.1001/archopht.1992.01080180084033.
26. Cheung, C.Y.; Thomas, G.N.; Tay, W.; Ikram, M.K.; Hsu, W.; Lee, M.L.; Lau, Q.P.; Wong, T.Y. Retinal Vascular Fractal Dimension and Its Relationship With Cardiovascular and Ocular Risk Factors. *Am. J. Ophthalmol.* **2012**, *154*, 663–674, doi:10.1016/j.ajo.2012.04.016.
27. Al-Sheikh, M.; Phasukkijwatana, N.; Dolz-Marco, R.; Rahimi, M.; Iafe, N.A.; Freund, K.B.; Sadda, S.R.; Sarraf, D. Quantitative OCT Angiography of the Retinal Microvasculature and the Choriocapillaris in Myopic Eyes. *Invest. Ophthalmol. Vis. Sci.* **2017**, *58*, 2063–2069, doi:10.1167/iovs.16-21289.
28. Azemin, M.Z.C.; Daud, N.M.; Ab Hamid, F.; Zahari, I.; Sapuan, A.H. Influence of Refractive Condition on Retinal Vasculature Complexity in Younger Subjects. *ScientificWorldJournal* **2014**, *2014*, 783525, doi:10.1155/2014/783525.
29. Li, H.; Mitchell, P.; Liew, G.; Rochtchina, E.; Kifley, A.; Wong, T.Y.; Hsu, W.; Lee, M.L.; Zhang, Y.P.; Wang, J.J. Lens Opacity and Refractive Influences on the Measurement of Retinal Vascular Fractal Dimension. *Acta Ophthalmol. (Copenh.)* **2010**, *88*, e234–e240, doi:http://dx.doi.org/10.1111/j.1755-3768.2010.01975.x.

30. Li, M.; Yang, Y.; Jiang, H.; Gregori, G.; Roisman, L.; Zheng, F.; Ke, B.; Qu, D.; Wang, J. Retinal Microvascular Network and Microcirculation Assessments in High Myopia. *Am. J. Ophthalmol.* **2017**, *174*, 56–67, doi:10.1016/j.ajo.2016.10.018.
31. Yang, Y.; Wang, J.; Jiang, H.; Yang, X.; Feng, L.; Hu, L.; Wang, L.; Lu, F.; Shen, M. Retinal Microvasculature Alteration in High Myopia. *Invest Ophthalmol Vis Sci* **2016**, *57*, 6020–6030.
32. Kurniawan, E.D.; Cheung, N.; Cheung, C.Y.; Tay, W.T.; Saw, S.M.; Wong, T.Y. Elevated Blood Pressure Is Associated with Rarefaction of the Retinal Vasculature in Children. *Invest. Ophthalmol. Vis. Sci.* **2012**, *53*, 470–474, doi:10.1167/iovs.11-8835.
33. Xu, Q.; Sun, H.; Huang, X.; Qu, Y. Retinal Microvascular Metrics in Untreated Essential Hypertensives Using Optical Coherence Tomography Angiography. *GRAEFES Arch. Clin. Exp. Ophthalmol.* **2020**, doi:10.1007/s00417-020-04714-8.
34. Grauslund, J.; Green, A.; Kawasaki, R.; Hodgson, L.; Sjolie, A.K. Retinal Vascular Fractals and Microvascular and Macrovascular Complications in Type 1 Diabetes. *Ophthalmology* **2010**, *117*, 1400–1405, doi:http://dx.doi.org/10.1016/j.ophtha.2009.10.047.
35. Kostic, M.; Bates, N.M.; Milosevic, N.T.; Tian, J.; Smiddy, W.E.; Lee, W.-H.; Somfai, G.M.; Feuer, W.J.; Shiffman, J.C.; Kuriyan, A.E.; et al. Investigating the Fractal Dimension of the Foveal Microvasculature in Relation to the Morphology of the Foveal Avascular Zone and to the Macular Circulation in Patients With Type 2 Diabetes Mellitus. *Front. Physiol.* **2018**, *9*, 1233, doi:https://dx.doi.org/10.3389/fphys.2018.01233.
36. Frydkjaer-Olsen, U.; Soegaard Hansen, R.; Pedersen, K.; Peto, T.; Grauslund, J. Retinal Vascular Fractals Correlate With Early Neurodegeneration in Patients With Type 2 Diabetes Mellitus. *Invest. Ophthalmol. Vis. Sci.* **2015**, *56*, 7438–7443, doi:10.1167/iovs.15-17449.
37. Tang, F.Y.; Ng, D.S.; Lam, A.; Luk, F.; Wong, R.; Chan, C.; Mohamed, S.; Fong, A.; Lok, J.; Tso, T.; et al. Determinants of Quantitative Optical Coherence Tomography Angiography Metrics in Patients with Diabetes. *Sci. Rep.* **2017**, *7*, doi:10.1038/s41598-017-02767-0.
38. Tang, F.Y.; Chan, E.O.; Sun, Z.; Wong, R.; Lok, J.; Szeto, S.; Chan, J.C.; Lam, A.; Tham, C.C.; Ng, D.S.; et al. Clinically Relevant Factors Associated with Quantitative Optical Coherence Tomography Angiography Metrics in Deep Capillary Plexus in Patients with Diabetes. *EYE Vis.* **2020**, *7*, doi:10.1186/s40662-019-0173-y.
39. Avakian, A.; Kalina, R.E.; Sage, E.H.; Rambhia, A.H.; Elliott, K.E.; Chuang, E.L.; Clark, J.I.; Hwang, J.-N.; Parsons-Wingerter, P. Fractal Analysis of Region-Based Vascular Change in the Normal and Non-Proliferative Diabetic Retina. *Curr. Eye Res.* **2002**, *24*, 274–280, doi:10.1076/ceyr.24.4.274.8411.
40. Zahid, S.; Dolz-Marco, R.; Freund, K.B.; Balaratnasingam, C.; Dansingani, K.; Gilani, F.; Mehta, N.; Young, E.; Klifto, M.R.; Chae, B.; et al. Fractal Dimensional Analysis of Optical Coherence Tomography Angiography in Eyes With Diabetic Retinopathy. *Invest Ophthalmol Vis Sci* **2016**, *57*, 4940–4947.
41. Bhardwaj, S.; Tsui, E.; Zahid, S.; Young, E.; Mehta, N.; Agemy, S.; Garcia, P.; Rosen, R.B.; Young, J.A. Value Of Fractal Analysis Of Optical Coherence Tomography Angiography In Various Stages Of Diabetic Retinopathy. *Retina Phila. Pa* **2018**, *38*, 1816–1823, doi:https://dx.doi.org/10.1097/IAE.0000000000001774.
42. Hirano, T.; Kitahara, J.; Toriyama, Y.; Kasamatsu, H.; Murata, T.; Sadda, S. Quantifying Vascular Density and Morphology Using Different Swept-Source Optical Coherence Tomography

Angiographic Scan Patterns in Diabetic Retinopathy. *Br. J. Ophthalmol.* **2019**, *103*, 216–221, doi:10.1136/bjophthalmol-2018-311942.

43. Cheung, N.; Donaghue, K.C.; Liew, G.; Rogers, S.L.; Wang, J.J.; Lim, S.-W.; Jenkins, A.J.; Hsu, W.; Li Lee, M.; Wong, T.Y. Quantitative Assessment of Early Diabetic Retinopathy Using Fractal Analysis. *Diabetes Care* **2009**, *32*, 106–110, doi:10.2337/dc08-1233.

44. Cheung, C.Y.; Sabanayagam, C.; Law, A.K.; Kumari, N.; Ting, D.S.; Tan, G.; Mitchell, P.; Cheng, C.Y.; Wong, T.Y. Retinal Vascular Geometry and 6 Year Incidence and Progression of Diabetic Retinopathy. *Diabetologia* **2017**, *60*, 1770–1781.

45. Ting, D.S.W.; Tan, G.S.W.; Agrawal, R.; Yanagi, Y.; Sie, N.M.; Wong, C.W.; San Yeo, I.Y.; Lee, S.Y.; Cheung, C.M.G.; Wong, T.Y. Optical Coherence Tomographic Angiography in Type 2 Diabetes and Diabetic Retinopathy. *JAMA Ophthalmol* **2017**, *135*, 306–312.

46. Kunicki, A.C.B.; Oliveira, A.J.; Mendonca, M.B.M.; Barbosa, C.T.F.; Nogueira, R.A. Can the Fractal Dimension Be Applied for the Early Diagnosis of Non-Proliferative Diabetic Retinopathy?. *Braz. J. Med. Biol. Res.* **2009**, *42*, 930–934, doi:http://dx.doi.org/10.1590/S0100-879X2009005000020.

47. Lim, S.W.; Cheung, N.; Wang, J.J.; Donaghue, K.C.; Liew, G.; Islam, F.M.A.; Jenkins, A.J.; Wong, T.Y. Retinal Vascular Fractal Dimension and Risk of Early Diabetic Retinopathy: A Prospective Study of Children and Adolescents with Type 1 Diabetes. *Diabetes Care* **2009**, *32*, 2081–2083, doi:10.2337/dc09-0719.

48. Koulisis, N.; Kim, A.Y.; Chu, Z.; Shahidzadeh, A.; Burkemper, B.; De Koo, L.C.O.; Moshfeghi, A.A.; Ameri, H.; Puliafito, C.A.; Isozaki, V.L.; et al. Quantitative Microvascular Analysis of Retinal Venous Occlusions by Spectral Domain Optical Coherence Tomography Angiography. *PLoS ONE* **2017**, *12*, doi:10.1371/journal.pone.0176404.

49. Remond, P.; Aptel, F.; Cunnac, P.; Labarere, J.; Palombi, K.; Pepin, J.L.; Pollet-Villard, F.; Hogg, S.; Wang, R.; MacGillivray, T.; et al. Retinal Vessel Phenotype in Patients with Nonarteritic Anterior Ischemic Optic Neuropathy. *Am J Ophthalmol* **2019**, *208*, 178–184.

50. Tzaridis, S.; Wintergerst, M.W.M.; Mai, C.; Heeren, T.F.C.; Holz, F.G.; Charbel Issa, P.; Herrmann, P. Quantification of Retinal and Choriocapillaris Perfusion in Different Stages of Macular Telangiectasia Type 2. *Invest. Ophthalmol. Vis. Sci.* **2019**, *60*, 3556–3562, doi:10.1167/iovs.19-27055.

51. de Mendonça, M.B. de M.; de Amorim Garcia, C.A.; Nogueira, R. de A.; Gomes, M.A.F.; Valença, M.M.; Oréfice, F. Fractal analysis of retinal vascular tree: segmentation and estimation methods. *Arq. Bras. Oftalmol.* **2007**, *70*, 413–422, doi:10.1590/s0004-27492007000300006.

52. Wainwright, A.; Liew, G.; Burlutsky, G.; Rochtchina, E.; Zhang, Y.; Hsu, W.; Lee, J.; Wong, T.-Y.; Mitchell, P.; Wang, J. Effect of Image Quality, Color, and Format on the Measurement of Retinal Vascular Fractal Dimension. *Invest. Ophthalmol. Vis. Sci.* **2010**, *51*, 5525–9, doi:10.1167/iovs.09-4129.

53. Lemmens, S.; Devulder, A.; Van Keer, K.; Bierkens, J.; De Boever, P.; Stalmans, I. Systematic Review on Fractal Dimension of the Retinal Vasculature in Neurodegeneration and Stroke: Assessment of a Potential Biomarker. *Front. Neurosci.* **2020**, *14*, 16, doi:10.3389/fnins.2020.00016.